\documentclass[fleqn,usenatbib]{mnras}

\usepackage{newtxtext,newtxmath}

\usepackage[T1]{fontenc}
\usepackage{ae,aecompl}

\usepackage{graphicx}	
\usepackage{amsmath}	
\usepackage{multicol}        
\usepackage{bm}		
\usepackage{pdflscape}	
\usepackage{url}

\title[Chrono-chemokinematics of the Galactic disc]{Unveiling the Distinct Formation Pathways of the Inner and Outer Discs of the Milky Way with Bayesian Machine Learning}

\author[I. Ciuc\u{a} et al.]{
Ioana Ciuc\u{a}$^{1}$,\thanks{E-mail: ioana.ciuca.16@ucl.ac.uk}
Daisuke Kawata$^{1}$,
Andrea Miglio$^{2}$,
Guy R. Davies$^{2}$,  
\newauthor{Robert J.J. Grand$^{3}$}
\\
$^{1}$Mullard Space Science Laboratory, University College London,	Holmbury St. Mary, Dorking, Surrey, RH5 6NT, UK \\
$^{2}$ Department of Physics, University of Birmingham\\
$^{3}$ Max-Planck-Institut f\"{u}r Astrophysik, Karl-Schwarzschild-Str. 1, 85748 Garching, Germany
}

\date{Accepted XXX. Received YYY; in original form ZZZ}

\pubyear{2020}

\begin{document}
\label{firstpage}
\pagerange{\pageref{firstpage}--\pageref{lastpage}}
\maketitle

\begin{abstract}
We develop a Bayesian Machine Learning framework called BINGO (Bayesian INference for Galactic archaeOlogy) centred around a Bayesian neural network. After being trained on the APOGEE and \emph{Kepler} asteroseismic age data, BINGO is used to obtain precise relative stellar age estimates with uncertainties for the APOGEE stars. We carefully construct a training set to minimise bias and apply BINGO to a stellar population that is similar to our training set. We then select the 17,305 stars with ages from BINGO and reliable kinematic properties obtained from \textit{Gaia} DR2.  By combining the age and chemo-kinematical information, we dissect the Galactic disc stars into three components, namely, the thick disc (old, high-[$\alpha$/Fe], [$\alpha$/Fe] $\gtrsim$ 0.12), the thin disc (young, low-[$\alpha$/Fe]) and the Bridge, which is a region between the thick and thin discs. Our results indicate that the thick disc formed at an early epoch only in the inner region, and the inner disc smoothly transforms to the thin disc. We found that the outer disc follows a different chemical evolution pathway from the inner disc. The outer metal-poor stars only start forming after the compact thick disc phase has completed and the star-forming gas disc extended outwardly with metal-poor gas accretion. We found that in the Bridge region the range of [Fe/H] becomes wider with decreasing age, which suggests that the Bridge region corresponds to the transition phase from the smaller chemically well-mixed thick to a larger thin disc with a metallicity gradient.   
\end{abstract}




\section{Introduction}
The Galactic disc is traditionally separated into the geometric thick and thin disc after \cite{Gilmore_1983} found from star counts that the vertical density profile of the Milky Way was better characterised by a superposition of two exponential profiles rather than one. High-resolution spectroscopic studies of the solar neighbourhood revealed also a bimodality in the chemistry of the disc, with the [$\alpha$/Fe]-[Fe/H] distribution showing distinct high- and low-[$\alpha$/Fe] components and a less prominent intermediate region \citep[e.g.,][]{Fuhrmann_1998, Pro_2000}. Beyond the local disc, recent large-scale spectroscopic surveys, such as the Apache Point Observatory Galactic Evolution Experiment (APOGEE), confirmed the existence of a similar high-[$\alpha$/Fe] sequence spanning a large radial and vertical extent of the Milky Way disc \citep[e.g.,][]{Anders_2014, Nidever_2014, Hayden_2015, Queiroz_2019}. The high-[$\alpha$/Fe] disc also appears to be thicker and more centrally concentrated than its low-[$\alpha$/Fe] counterpart \citep[e.g.,][]{Bensby_2011, Bovy_2012, Cheng_2012}.

One of the first approaches to explain the chemical bimodality seen in the Galactic disc is the two-infall model, a numerical chemical evolution model developed by \cite{Chiappini_1997}, \cite{Chiappini_2001}, \cite{Grisoni_2017}, \cite{Spitoni_2019}. \cite{Chiappini_2001} suggested that the high-[$\alpha$/Fe], chemically homogenous disc forms early during an intense star formation period dominated by Type II supernovae (SNe II) following a rapid infall of primordial gas. After a brief cessation in star formation, the second episode of gas accretion takes place that lowers the metal content in the interstellar medium due to the continuous infall of low metallicity fresh gas. The low-[$\alpha$/Fe] disc then builds up gradually from lower [Fe/H]. \cite{Bekki_2011} also follow a semi-analytical approach to explain the existence of two distinct populations. In their continuous star formation model, the high-[$\alpha$/Fe] sequence up to around solar [Fe/H], i.e. the thick disc, forms early during a rapid, intense star formation period. The thin disc then proceeds to form gradually from the remaining gas with solar [Fe/H] and [$\alpha$/Fe]  mixed with the fresh primordial gas accreted after the formation of the thick disc.A sequence of increasing [Fe/H] and decreasing [$\alpha$/Fe] builds up gradually. Still, this sequence is lower in [$\alpha$/Fe] as Type Ia SNe can already enrich the environment at this time. Once star formation reaches its peak and starts decreasing, a sequence with decreasing [$\alpha$/Fe] and increasing in [Fe/H] follows along with the same low-[$\alpha$/Fe] sequence.

More recent scenarios inspired by galactic dynamics proposed that radial migration of kinematically hot stars formed in the inner disc builds up a thick disc after moving outward in the disc \citep{Scho_2009, Loebman_2011, Roskar_2013}. Radial migration is successful in explaining the age-metallicity and metallicity-rotation velocity relation observed in the Milky Way. However, there is still considerable debate regarding the efficiency of radial migration in building a geometrically thick disc \citep[e.g.,][]{Minchev_2012, Grand_2016, Kawata_2017}.

High-resolution numerical simulations also suggested several thick and thin disc formation scenarios, including violent gas-rich mergers at high-redshift \citep[e.g.,][]{Brook_2004, Grand2018, Grand_2020}, accretion of high-[$\alpha$/Fe] stars \citep{Abadi_2003, Kobayashi_2011, Tissera_2012}, vertical heating from satellite merging events \citep[e.g.,][]{Quinn_1993, Villalobos_2008} and turbulence in clumpy high-redshift gas-rich disc \citep{Noguchi_1998, Bournaud_2009, Beraldo_2019}. The recent popular view is that the thick disc formation precedes the thin disc formation and the earlier disc was smaller and thicker, i.e. an inside-out and upside-down formation of the disc \citep[e.g.,][]{Brook_2004, Brook_2006, Bird2013}. In \cite{Brook2012}, the majority of the thick-disc stars form as gas originating from a gas-rich merger at high-redshift settles into a disc at the end of the merger epoch. This early disc is kinematically hot and radially compact. Once the chaotic phase of the star formation of the thick disk ends, the younger, lower [$\alpha$/Fe] thin disc can gradually grow in an inside-out fashion \citep[e.g.,][]{Matteucci_1989} as gas is smoothly accreting to the central galaxy. As in \cite{Brook2012}, \cite{Noguchi_2018} and \cite{Grand2018} suggested that chemical evolution proceeds at different rates in the inner and outer disc, resulting in more chemically evolved stars in the inner regions.   Radial migration can bring the thick disk stars formed in the inner disc to the outer disc at redshift $z \sim 0$, so that we can observe the thick disc stars at the solar neighbourhood \citep{Brook2012, Minchev_2013}.

\begin{figure}
	\centering
	\includegraphics[width=\columnwidth]{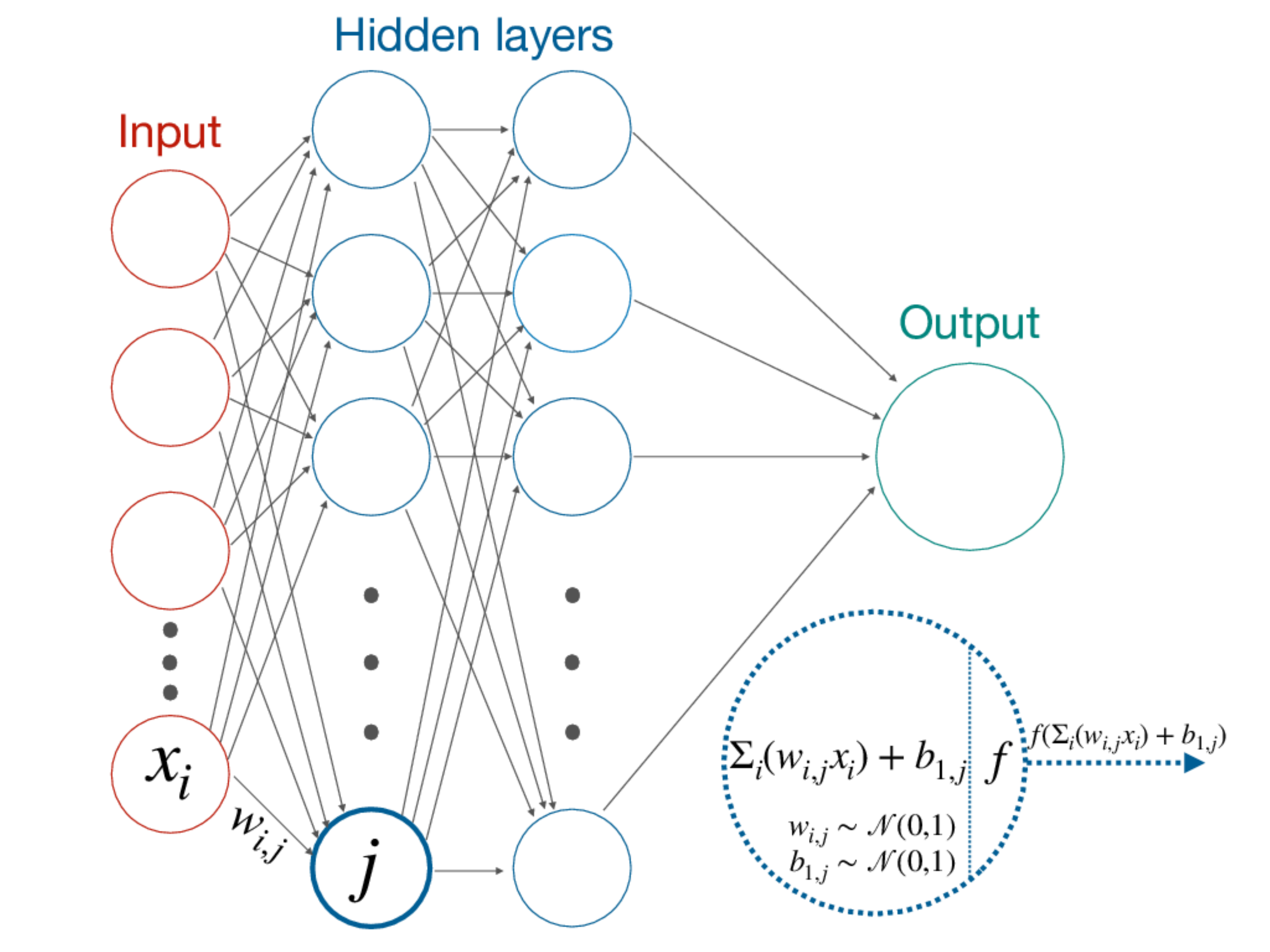}
	\caption{Schematics of a Bayesian neural network with 2 hidden layers. Each connection between neurons has an associated weight and the neurons in the the hidden and output layers have an associated bias. The connection between neurons $i$ in the input layer and $j$ in the first hidden layer has the associated weight $w_{i,j}$ and the neuron $j$ has a bias $b_{1,j}$. Each weight and bias parameters have an associated prior $\mathcal{N}(0, 1)$. The dotted circle is a zoom-in of the neuron $j$, and shows the transformation applied to the input data $x_i$ in the first hidden layer of the network, namely $x_i \rightarrow f(\Sigma_{i}(w_{i,j} x_i) + b_{1,j}
)$, where $f$ is the activation function. We use a rectified linear unit (ReLU) activation function in our analysis.}
	\label{fig:schematics}
\end{figure}

\begin{figure}
	\centering
	\includegraphics[width=\columnwidth]{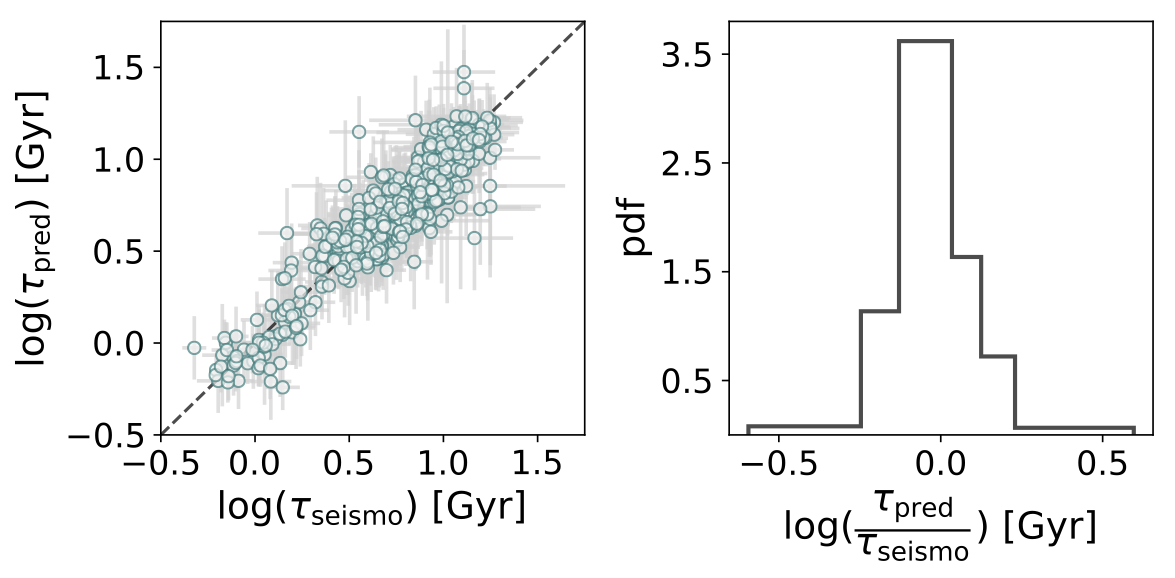}
	\caption{Comparison between observed (target) logarithmic age of stars, $\log(\tau_{\rm seismo})$, derived with asteroseismology and the predicted $\log(\tau_{\rm pred})$ by BINGO. The panels show the results when applying the model trained with the age data augmented training set with the distance shuffling (Model A) to the original test data (Test 1, see Sec. \ref{sec:trainingset} for details). The light green circles in the left panel show the model prediction results against the observed target age. The standard deviation in the prediction and observation are shown as the grey lines. The right panel shows the difference between prediction and target, which peaks at 0 with a standard deviation of 0.1~dex.}
	\label{fig:modelA_test1}
\end{figure}

The \textit{Gaia} mission \citep[DR2, ][]{Gaia2018} is providing information to obtain the accurate position and motion for more than a billion stars in the Milky Way. The APOKASC-2 catalogue \citep{Pins2018}, comprised of 6,676 evolved stars in the APOGEE DR14 survey observed by the \emph{Kepler} mission \citep{Bor2010}, provides the best asteroseismology information to infer the age for giant stars, which is crucial for Galactic archaeology. In this paper, we use a state-of-the-art machine learning method, a Bayesian neural network, trained on the APOKASC-2 data, to obtain reliable relative stellar age estimates for 17,305 carefully selected disc stars in the  APOGEE data. We use the age, chemistry and kinematical information to examine the formation history of the Galactic disc by comparing our results with what is expected from the formation scenarios of the thick and thin disc suggested by the recent numerical simulations described above.

This paper is organised as follows. Section~\ref{sec:BINGO} describes the Bayesian Machine Learning framework, called BINGO (Bayesian INference for Galactic archaeOlogy), that we employ in the current analysis. We discuss here how the biases in the training dataset affect the performance of the neural network model and our approach to minimise the bias in the subsequent inferences. In Section \ref{sec:results}, we present the results after applying BINGO to carefully selected stars in the APOGEE survey. A brief discussion of our results is given in Section \ref{sec:disc}. Finally, a summary of our findings is given in Section~\ref{sec:summary}.

\section{Method}
\label{sec:BINGO}
In this paper, we introduce BINGO  which is a Bayesian Machine Learning framework to obtain stellar ages of evolved stars using photometric information from the second data release of the European Space Agency's (ESA) \textit{Gaia} mission \citep[\textit{Gaia} DR2, ][]{Gaia2018} and the stellar parameter information from the fourteenth data release of the SDSS-IV APOGEE-2 \citep{Maj2017}. BINGO consists of a Bayesian neural network trained using the asteroseismic age determined from <$\Delta \nu$> from the individual radial-mode frequency from the \emph{Kepler} light curve \citep{Miglio_2020}.

\begin{figure*}
	\centering
	\includegraphics[width=\textwidth]{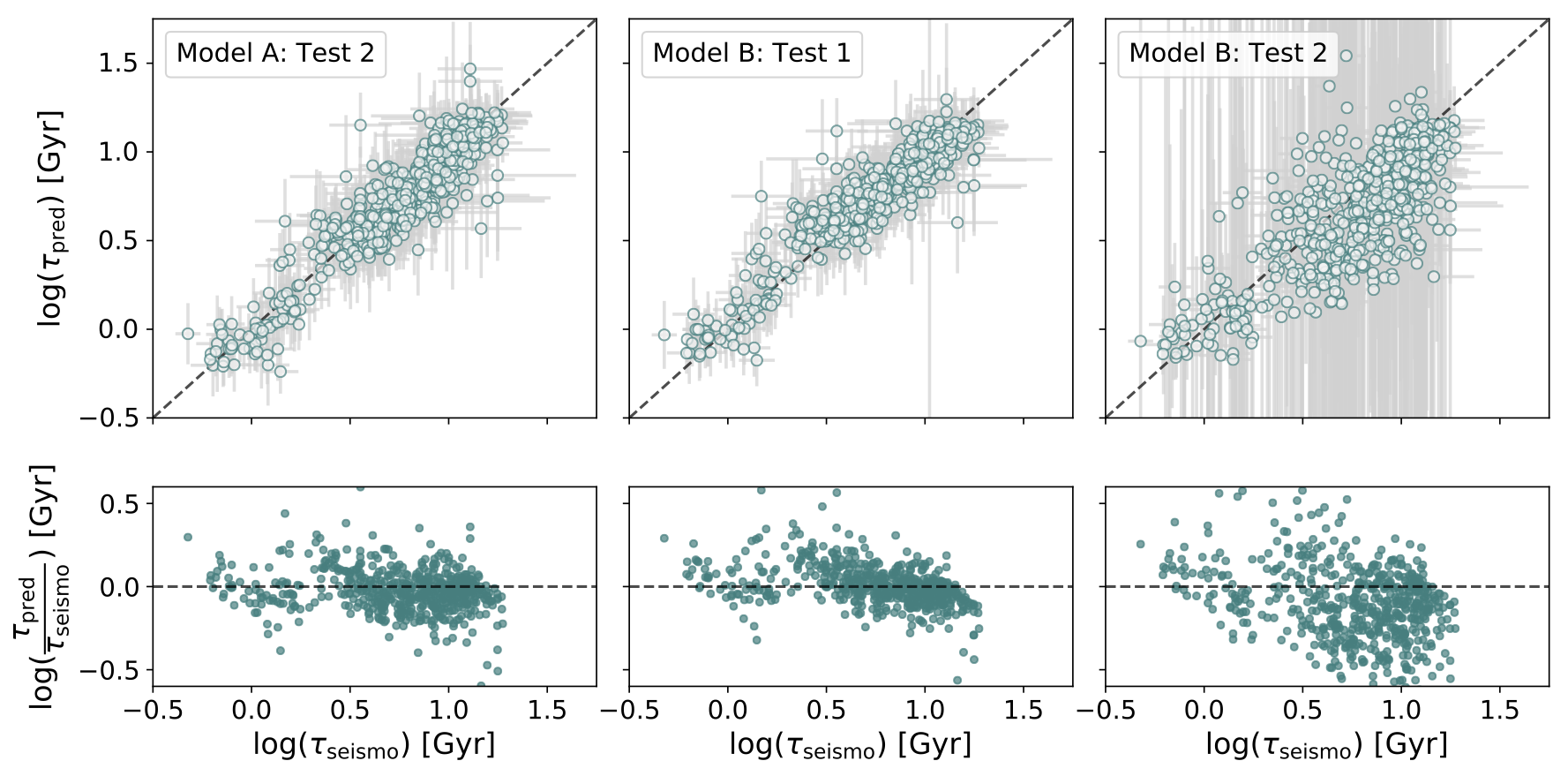}
	\caption{Predictions vs the target asteroseismic age in $\log(\tau)$. Left panel shows the result from a model trained on the age data augmented training set with the distance shuffling (Model A) and applied to the distance shuffled test set (Test 2). The middle and right panel show the predictions for a model trained on the original data (Model B) applied to the original (Test 1) and distance shuffled test data (Test 2), respectively. The standard deviation in the prediction and asteroseismic age are shown as the grey lines. Model A predictions for Test 2 performs better than Model B prediction for Test 1. The model trained on the original data and applied to the distance shuffle data, i.e. Model B prediction for Test 2, performs considerably worse as Model B has learned the distance dependence of age and metallicity, which is erased in Test 2.}
	\label{fig:comparison}
\end{figure*}

\textit{Gaia} DR2 provides astrometric information to obtain the position and proper motion for $\sim$ 1.3 billion stars with unprecedented accuracy \citep{Lind2018} as well as high-quality multi-band photometry for a large subset of these stars \citep{Riello2018, Evans2018}. For a selected type of stars with a G-band magnitude between about 4 and 13 magnitudes, the mean line-of-sight velocities measured with \textit{Gaia} Radial Velocity Spectrometer (RVS), line-of-sight velocities have also been provided in \textit{Gaia} DR2 \citep{cropper2018gaia, Sartoretti_2018, Katz_2019}. We use the photometric data from \textit{Gaia} DR2 for BINGO, and the parallax and proper motion information to derive the kinematic properties for our sample of stars.

APOGEE is a spectroscopic survey in the near-infrared H-band (15,200 \AA-16,900 \AA) with a high resolution of R $\sim$ 22,500, observing more than 200,000 stars (as of DR14) located primarily in the disc and bulge of the Milky Way. In this work, we employ the calibrated stellar parameters such as effective temperature and surface gravity as well as metal abundances obtained with the APOGEE Stellar Parameters and Chemical Abundances Pipeline \citep[ASCAP, ][]{Gar2016} in the APOGEE DR14 survey \citep{Abol2018}. In addition, we use the 2MASS J, H, K photometry and their associated uncertainties \citep{Skru2006} reported in the APOGEE DR14 catalogue.

\subsection{BINGO}
Machine Learning has revolutionised the way we perform data analysis tasks in Astronomy, which has grown into a big data field with the emergence of large surveys such as SDSS and \textit{Gaia}. Neural networks are Machine Learning methods that can, in principle, model any smooth map between a high-dimensional input data to a set of desirable outputs. Depending on their architecture, neural networks can consist of one or more fully-connected layers, each with a number of neurons that essentially take the input and transform it through linear activation functions to an output of interest (also known as feed-forward artificial neural networks). In supervised learning, which BINGO uses, the parameters of the neural network, e.g. weights that define the connection between neurons, are trained and optimised to best reproduce the training set where the input and output are known. Then, the trained neural network can be applied to the data whose output is unknown with much less computational cost than training. 

In Bayesian Inference, the power of Bayes' Law is that it allows us to relate the probability of a model given the data to a quantity that is easier to understand, namely the probability we would observe the data given the model and any background information, $I$, i.e.,
\begin{equation}
p(model|data,I) \propto p(data|model, I)p(model|I),
\label{eq:bayes}
\end{equation}
where $p(model|data, I)$ is the posterior probability, $p(data|model, I)$ is the likelihood and $p(model|I)$ is the prior. The posterior encompasses our state of knowledge about a model given that we gather new data through the likelihood. Following equation (\ref{eq:bayes}), Bayes' Law can be applied to a neural network to come up with a probability distribution over its model parameters\footnote{\url{https://twiecki.io/blog/2016/07/05/bayesian-deep-learning/}} and construct a Bayesian Neural Network as done in the pioneering work of \cite{Das_2018} and \cite{Sanders_2018}. This powerful synergy between Bayesian Inference and Machine Learning allows us to naturally introduce uncertainty into our machine learning approach, i.e. we can get an estimate of how confident our neural network is of its predictions. 

BINGO's architecture consists of 2 fully connected layers with 16 neurons each (Fig.~\ref{fig:schematics}). We use the probabilistic programming framework {\tt pymc3} \citep{Salv_2015} and its Magic Inference Button, the No-U-Turn-Sampler (NUTS) as the MCMC sampler. We use a Gaussian prior of $\mathcal{N}(0, \sigma)$ for the weights and bias parameters in the neural network, which effectively acts as L2 regularisation. It is possible to optimise $\sigma $ of the Gaussian prior, but it is computationally too expensive. Therefore, we adopt $\sigma=1$ for simplicity. We use 4 chains that allow us to diagnose our samples and make sure the samples returned from the NUTS sampler are drawn from the target distribution. Once we have a posterior distribution over the neural network parameters, we can then compute a distribution over the network outputs by marginalising over the network parameters. We note that this Bayesian Neural Network scheme assumes that all the input features, such as the stellar parameters, are independent, and cannot take into account the covariance between the inputs. It is also worth noting that the neural network model depicted in Fig.~\ref{fig:schematics} is not identifiable \citep{Pourzanjani_2017}. Hence, the naive MCMC sampling of the network parameters suffers from the unidentifiability of the parameters. Still, we have confirmed that the posterior distribution of the target age prediction from the 4 different chains are consistent with each other. Therefore, we are confident that our age prediction, especially the mean of the prediction used in this paper, does not suffer severely from unidentifiability. These known challenges for Bayesian Neural Networks remain caveats of BINGO, upon which we hope to improve in a future study. 

\subsection{Building an effective training set}
\label{sec:trainingset}
In this study, we employ a training set created from the APOKASC-2 dataset with our derived asteroseismic age \citep{Miglio_2020}. The assumptions used to derive the asteroseismic ages that we are using are given as R11 in Table 1 of \cite{Miglio_2020}. We select only red clump stars (RC) with masses higher than 1.8 $M_{\odot}$ and the red giant branch (RGB) stars, for which the relative asteroseismic ages are reliable. To construct our base training set, we use only stars with high signal-to-noise (SNR) APOGEE spectra (SNR > 100), which leaves us with 2,915 stars.  We then use the APOGEE stellar parameters and photometry data, T$_{\rm eff}$, $\log$ $g$, [$\alpha$/M], [M/H] \footnote{In APOGEE DR14, $\alpha$-elements comprise of O, Mg, Si, S, Ca and Ti. We used the ASPCAP measurements of [$\alpha$/M] and [M/H] as a proxy for [$\alpha$/Fe] and [Fe/H], respectively. Correspondingly, we use the labels of  [$\alpha$/Fe] and [Fe/H] to refer to [$\alpha$/M] and [M/H].}, [C/Fe], [N/Fe], G, BP, RP, J, H and K as the input features in BINGO to map them to the common logarithm of the asteroseismic age, $log(\tau)$, referred to as the target. 

Because the original data comes from a limited \emph{Kepler} field data, our original training set has a known dependence of age and metallicity on the distance (which affects photometry). Also, there are not many young or old stars in our selected RGB and RC data. To correct for the distance dependence, we randomly displace the distance of stars between 0 and 10~kpc and then adjust the apparent magnitude of the stars depending on the difference between the new distance and the original distance. We do not change the extinction upon displacing the distance also to erase the dependence of extinction on the distance. We refer to this technique as distance shuffling. 

Our training set contains a smaller number of young (age $< 2$~Gyr) stars and very old (age > 12~Gyr) stars, and this imbalance becomes more apparent when using $\log(\tau)$ as our target variable for BINGO. During training, the model learns to reproduce the target variable only for a majority of intermediate age stars, which biases the prediction toward the intermediate age irrespective of their true age, and consequently leads to overpredicting the age of younger stars and underpredicting the age of the very old stars, an effect also known as regression dilution. To minimize the effect of this bias and balance our training set, we effectively oversample the fewer young stars and very old stars to balance the number of stars at different $\log(\tau)$. To this end, we first examine the distribution of our original training set in $\log(\tau)$. We then use a Kernel Density Estimator (KDE) to approximate the distribution in $\log(\tau)$, and for each star, we find its probability under the KDE, which we refer to as $prob$. We then compute the inverse probability and round it the nearest integer, N = [1/$prob$]. Following the distance shuffling procedure described above, we randomly distance-shuffle each star N times. This approach leads to some of the stars in the original dataset to be sampled more than once.  Since their distances and hence their apparent magnitude are different,  these ``artificial" stars become members of an augmented training set. Since we are using data augmentation, which is an established machine learning technique, we refer to our approach as age data augmentation. The final training set has 4,673 stars after performing the age data augmentation technique on the training set data (80\% of the original data). Note that the data augmentation can reduce the uncertainties in our predictions, because we artificially increase the number of data points. Therefore, our uncertainties do not statistically reflect the uncertainty in the measurement of the stellar age. In this paper, however, our priority is to mitigate regression dilution with this simple data augmentation technique. This is another caveat of BINGO in addition to the assumed independence of the input features and the unidentifiability discussed above. In this paper, as described later, we use the uncertainties only as the metric of confidence of our prediction, and do not use the uncertainties for any quantitative discussion. Hence, the discussion of this paper is unlikely to be affected by these issues. We postpone the resolution of these issues to a future study.

To evaluate the prediction accuracy of BINGO, we split our original data of 2,915 stars into training (80 \%, 2,331 stars) and testing (20 \%, 583 stars) data. To demonstrate the importance of distance shuffling and age data augmentation, we consider two different trained models: Model A trained on the age data augmented training set of 4,673 stars with the distance shuffling and Model B trained on the original training set without the distance shuffling or the age data augmentation. Then, we create a testing set, Test 1, which is the 20\% of the original data which are not used for training, and Test 2 which is the same data as Test 1, but the distance has been shuffled. Fig. \ref{fig:modelA_test1} shows the predictions from Model A on Test 1 and the error associated with the prediction. The asteroseismic age is well reproduced by the prediction from BINGO Model A, with a standard deviation  $\sim$ 0.1~dex. Note that the ages of some of old stars are much older than the age of the Universe. This is because there is no prior of the maximum age considered in our asteroseismic age measurement \citep{Miglio_2020}.

Fig. \ref{fig:comparison} presents the predictions from Model A on Test 2 (left), from Model B on Test 1 (middle) and Test 2 (right). There is little difference between Model A on Test 1 (see Fig. \ref{fig:modelA_test1} and Model A on Test 2. This means that BINGO Model A can recover the age well in application data which have no distance dependence in age or metallicity. The middle panel of Fig. 2 shows that Model B trained on the original dataset without the age data augmentation leads to a systematic overprediction for the age of stars with the asteroseismic age of $\log(\tau_{\rm seismo})<0.5$~dex and underprediction of the age for stars with $\log(\tau_{\rm seismo}$) > 1.0~dex. This is because Model B is trained mainly to reproduce the overwhelming number of stars with 0.5 < $\log(\tau_{\rm seismo})$ < 1.0~dex and suffers from the regression dilution effect mentioned above. The right panel of Fig. \ref{fig:comparison} shows the age prediction of Model B on Test 2, which shows much worse recovery of the asteroseismic ages with large uncertainties. This is because Model B has learned the dependence of the age and metallicity on the distance in the original training set. These results demonstrate why it is important to erase the distance dependence in the training set and keep the balance of the number of sample in the output label, i.e., $\log(\tau)$. We therefore use Model A in this paper. 

\subsection{RGB and high mass RC selection}
\label{sec:rgbrcsel}

Our training set consists of the specific population of RC stars with a mass higher than $1.8 \ M_{\odot}$ and RGB stars in the limited \emph{Kepler} field. When we apply our trained model to the rest of APOGEE data, we select only the same population as the population of the training data. 
Hence, we train a 3-layer artificial neural network on the original APOKASC-2 data to classify RC stars with a mass higher than $1.8 \ M_{\odot}$ and RGB stars. For this classification task, we train the model using {\tt Keras} and {\tt TensorFlow} \citep{Abadi_2016}, which is much less computationally expensive than training a Bayesian Neural Network. The selection function of APOKASC-2 is not the same as the rest of the APOGEE data. However, because we need the stellar mass and the RC, RGB and AGB classification for the training and validation data, we use the APOKASC-2 data for training and validation. We have constructed a classification neural network to identify the RC stars with > 1.8 $M_{\odot}$ and RGB stars using our asteroseismic analysis of the APOKASC-2 data. We used the input features of T$_{\rm eff}$, $\log$ $g$, [$\alpha$/Fe], [Fe/H], [C/Fe], [N/Fe], G, BP, RP, J, H and K, and 2,948 positive, i.e. the high mass RC or RGB, and 1,918 negative stars are used for the training. Similar strategies are employed in \cite{Hawkins_2018} and \cite{Ting_2018} to identify the RC stars.

\begin{figure}
	\centering
	\includegraphics[width=\columnwidth]{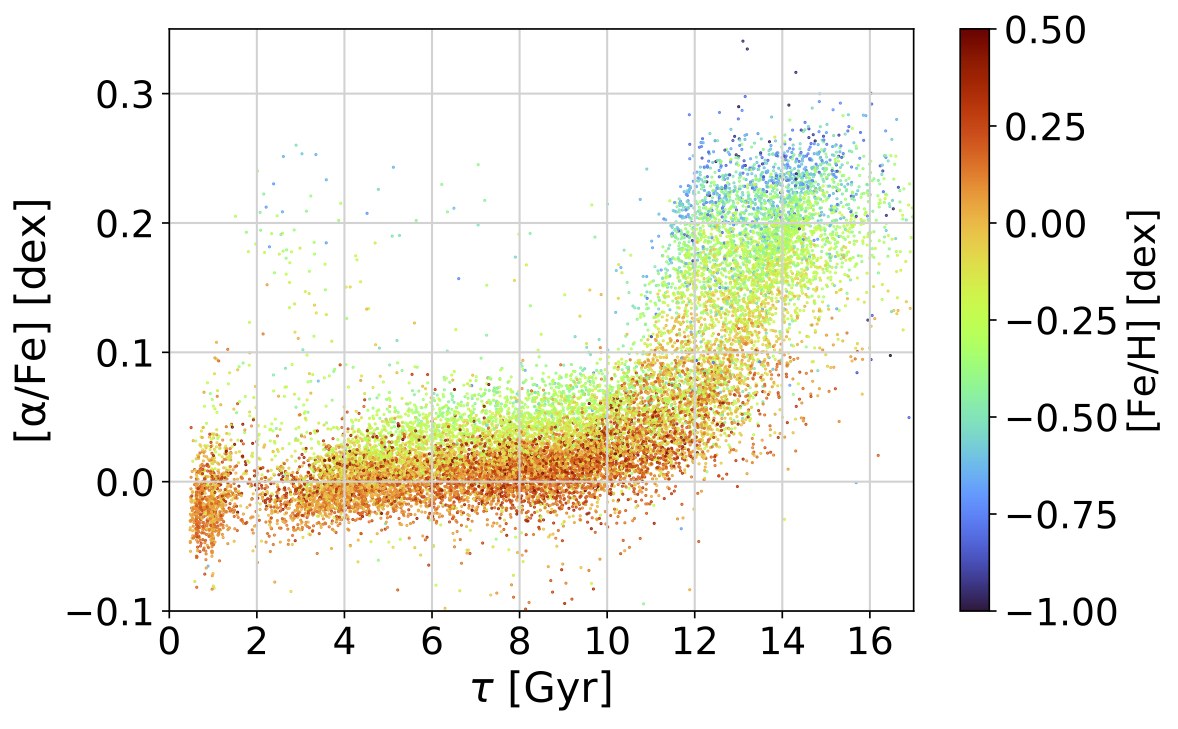}
	\caption{[$\alpha$/Fe] as a function of age coloured by metallicity, [Fe/H]. The high-[$\alpha$/Fe] population ([$\alpha$/Fe]$>0.1$~dex) is older and more metal-poor.}
	\label{fig:age-afe-mh}
\end{figure}

\begin{figure}
	\centering
	\includegraphics[width=\columnwidth]{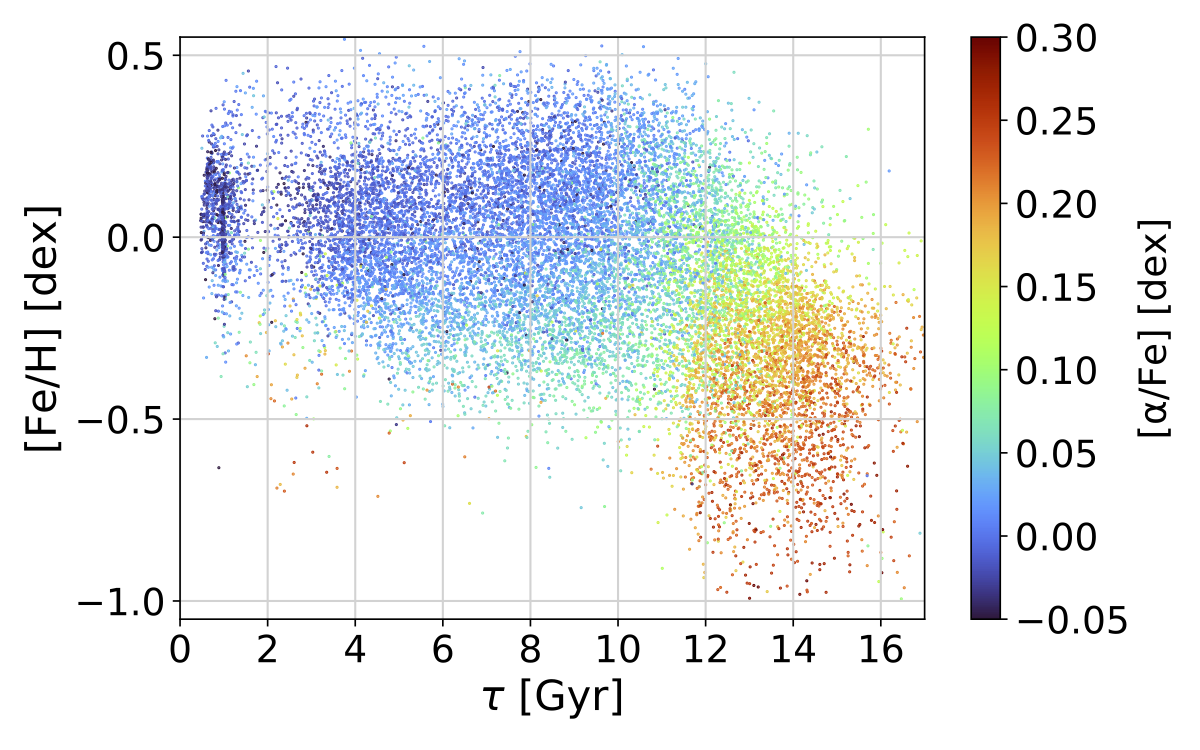}
	\caption{[Fe/H] as a function of age coloured by [$\alpha$/Fe]. The younger population is more metal-rich than its older counterpart. Stars more metal-poor than $-0.5$~dex are considerably old.}
	\label{fig:age-mh-alphafe}
\end{figure}

\begin{figure*}
	\centering
	\includegraphics[width=\textwidth]{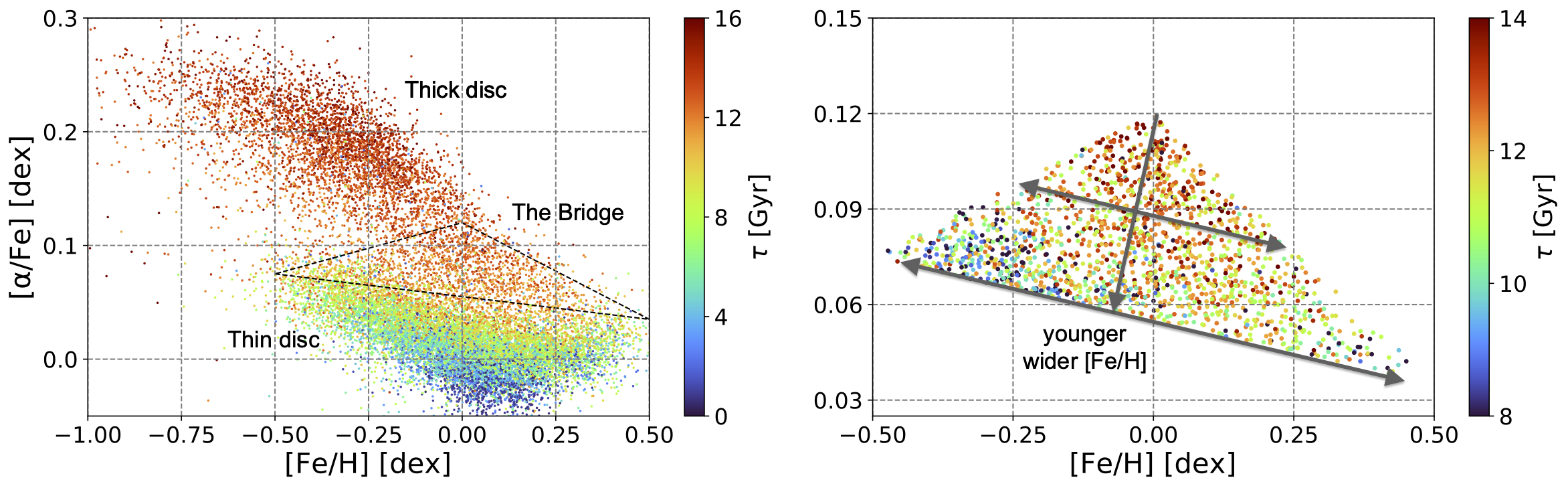}
	\caption{The distribution of [$\alpha$/Fe] and [Fe/H] coloured by age for our sample of stars. We refer to the old high- and young low-[$\alpha$/Fe] populations as the thick and thin disc, respectively, as highlighted in the left panel. The dotted triangle region in the left panel is referred to as the Bridge, and is a transition region between the thick (high-$\alpha$ sequence with [$\alpha$/Fe] $\gtrsim$ 0.1~dex) and thin disc (low-[$\alpha$/Fe] sequence). An age gradient is apparent, as indicated by the near vertical downward arrow, in a close-up of the Bridge region, shown in the right panel, and the range of [Fe/H] becomes wider, as indicated by the two near horizontal arrows for the younger stars in the Bridge region.
	}
	\label{fig:afe-feh-age}
\end{figure*}

We then use the trained neural network model to classify stars in the APOGEE cross-matched with \textit{Gaia} DR2 dataset \citep{Sanders_2018}. We also limit our data to having APOGEE spectra with SNR > 100 and the $K$-band extinction smaller than 0.1~mag in the APOGEE catalogue, because all of our training data has the $K$-band extinction $<0.1$~mag. We only select stars that have a probability higher than 95 \% of being classified as RC with higher mass than $1.8 \ M_{\odot}$ or RGB. We apply the BINGO Model A to this selected data to get the posterior probabilities for $\log(\tau)$. 

Our strategy in this paper is to use the most reliable data only. We therefore select stars with $\log (\tau)$ age uncertainties less than 10~\%. Note that the age uncertainties from BINGO indicate epistemic uncertainties of the model prediction, which can be smaller than the observed uncertainty of the original asteroseismic age. Also, to obtain reliable kinematic properties from the \textit{Gaia} data, we select the data with parallax uncertainties of $\pi / \sigma_{\pi} > 5.0$. We compute the distance using the \textit{Gaia} parallax with the additional systematic bias of parallax of 54 $\mu$as \citep{Schonrich_2019}, and select the stars in the limited volume of $7 < R < 9$~kpc and $z < 2$~kpc, where we assume the solar position at the Galactocentric distance of 8~kpc and the vertical height of the Sun from the disc plane of 0.025~kpc. We obtain kinematic properties using {\tt galpy} \citep{Bovy2015}. We have confirmed that our derived age and kinematics are consistent with \cite{Sanders_2018}, except for the difference in the absolute age scales, because we use a different asteroseismic age scale for our training set \citep{Miglio_2020}. As a result, we obtain 17,305 stars, which are used in the following sections.

\section{Results}
\label{sec:results}
In this section, we explore the relations between stellar age, chemistry and orbital properties for our sample of stars. Reliable relative age estimates for a large number of stars obtained with BINGO enable us to find that the inner and outer discs follow a different formation and chemical evolution pathway. Our results provide further evidence for an upside-down, inside-out formation of the Galactic disc.

\subsection{The chrono-chemical map of disc stars}
We first investigate the evolution of $\alpha$-abundances, [$\alpha$/Fe], and metallicity, [Fe/H], with age, $\tau$. Fig. \ref{fig:age-afe-mh} shows the enhancement in [$\alpha$/Fe] as a function of age coloured by metallicity. The deficiency of stars with age $\sim$ 1.5~Gyr arises because we select the RC stars with mass > 1.8 $M_{\odot}$ and there are considerably fewer RGB stars with ages younger than 3~Gyr. The high-[$\alpha$/Fe] ``sequence" separates clearly from the low-[$\alpha$/Fe] ``sequence" in the age-[$\alpha$/Fe] space at [$\alpha$/Fe] $\sim$ 0.1~dex, where there seems to be a population gap extending approximately 0.02~dex. The majority of the high-[$\alpha$/Fe] stars ([$\alpha$/Fe] > 0.1~dex) are generally older and more metal-poor than the low-[$\alpha$/Fe] population. [$\alpha$/Fe] rapidly decreases with decreasing age up to $\sim$  10~Gyr. The age-[$\alpha$/Fe] relationship also appears to be broader in [$\alpha$/Fe] at a fixed age for the high-[$\alpha$/Fe] population, in qualitative agreement with \cite{Aguirre_2018}.
However, considering the uncertainties of the age, this relationship is considered to be tight \citep{Haywood_2013,Snaith_15}. A striking feature of Fig. \ref{fig:age-afe-mh} is the young, low metallicity, high-$\alpha$ stars, also seen in \cite{Chiappini_2015}, \cite{Martig_2015}, \cite{Aguirre_2018}. We discuss the origin of this population in more detail in Section \ref{younghal}.

\begin{figure*} 
	\centering
	\includegraphics[width=\textwidth]{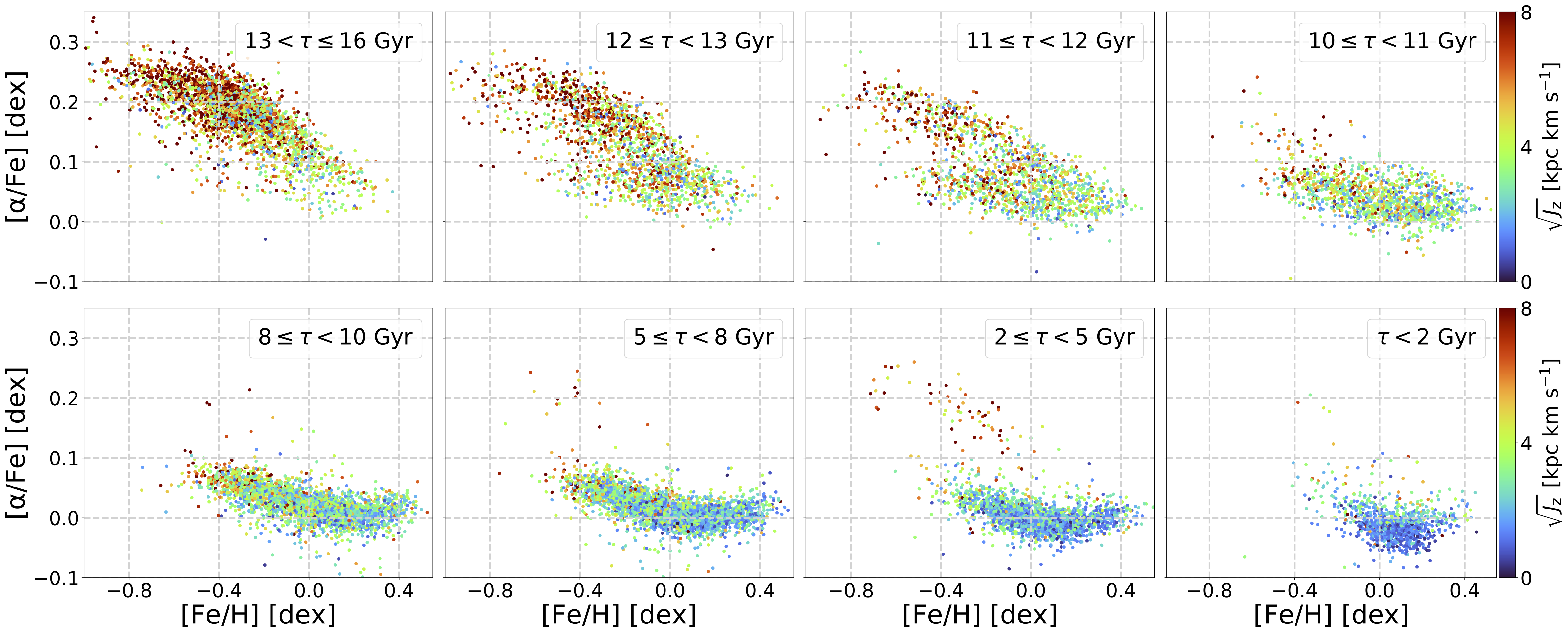}
	\caption{The distribution of [$\alpha$/Fe] and [Fe/H] coloured by the square root of the vertical action, $\sqrt{J_z}$, for the samples of stars within 8 different age bins. The top four panels show [Fe/H]-[$\alpha$/Fe] relationship for the older stars ($10 < \tau < 16$~Gyr), and the lower four panels present those for the younger stars ($\tau$ < 10~Gyr). There is a kinematically hot population of young high-[$\alpha$/Fe] stars seen in the lower panels.}
	\label{fig:al-feh-jz}
\end{figure*}

Fig. \ref{fig:age-mh-alphafe} shows the age-metallicity relationship coloured with [$\alpha$/Fe]. While the old, high-[$\alpha$/Fe] stars exhibit a clear trend of decreasing [Fe/H] with age, the younger, low-[$\alpha$/Fe] disc shows a flat age-[Fe/H] relation up to $\sim$ 11~Gyr. For stars with [Fe/H] $>-0.5$~dex, our results are qualitatively similar to those from previous studies, such as \cite{Cas2011}, \cite{Aguirre_2018} and \cite{Mackereth_2019}. For the metal-poor and high [$\alpha$/Fe] population, the tight trend observed between age and metallicity is consistent with \cite{Bensby_2005} and \cite{Haywood_2013}, who analysed dwarf stars and used the isochrone age.

In Fig. \ref{fig:afe-feh-age} we examine the distribution in [$\alpha$/Fe] and [Fe/H] coloured by age for the stars in our sample. Classically, this diagram is used to identify the chemically defined thick (high-[$\alpha$/Fe]) and thin (low-[$\alpha$/Fe]) disc stars. Using the age alongside the chemical information, we define three regions in the [$\alpha$/Fe]-[Fe/H] space, namely an old thick disc (high-[$\alpha$/Fe], low-[Fe/H]), young thin disc (low-[$\alpha$/Fe], broad [Fe/H]) and the Bridge (high-[$\alpha$/Fe], high-[Fe/H]). The Bridge stars\footnote{There are 1,682 stars in the Bridge, which represents approximately 10\% of the entire sample. Caution must be used when using this number as it can be sensitive to the selection function.} are selected to lie within the triangle region starting around ([Fe/H], [$\alpha$/Fe]) = (0.0, 0.12)~dex. \cite{Anders2018chem} suggested that the population of stars found in this region had a different origin and history to the thick and thin disc stars. Our results further suggest that the Bridge population appears to be a transition region connecting the old thick and the young thin disc. The right panel of Fig. \ref{fig:afe-feh-age} reveals a noticeable age gradient within this population from the oldest, more [$\alpha$/Fe]-enhanced and metal-rich stars to a younger population spanning a broader distribution of metallicities from [Fe/H] $=-0.5$ to 0.5~dex. Although this age gradient in the Bridge is tentative, we notice that a similar trend is also seen in \cite{DelgadoMena_2019} who studied 1,000 FGK dwarf stars from the HARPS-GTO programme and analysed the isocrhone ages of these stars. Therefore, it is reassuring that the trend shown in our asteroseismic-trained ages of giants is similar to that based on the isochrone ages of dwarfs.

\subsection{Chemo-kinematical analysis}
\label{dyn}

To connect the observed stellar chemical properties to kinematic properties, we compute the vertical action and the mean orbital radius for our sample of stars using {\tt galpy} in the fixed {\tt MWPotential2014} configuration \citep[version 1.5,][]{Bovy2015}. Although {\tt MWPotential2014} is time-independent and axisymmetric and cannot capture disequilibrium effects present in the real disc, it is nonetheless a good and useful approximation for the purpose of this work. Fig. \ref{fig:al-feh-jz} shows the distribution of [$\alpha$/Fe] and [Fe/H] as a function of age and vertical action $J_z$. The general trend is that $J_z$ is decreasing with age, with the older population being significantly hotter than the younger population. As also inferred from Fig. \ref{fig:afe-feh-age}, Fig. \ref{fig:al-feh-jz} clearly shows that the Bridge region starts appearing at age $< 13$~Gyr and it spreads to lower [$\alpha$/Fe] and to a wider range of [Fe/H] with decreasing age as seen in the triangle features at [$\alpha$/Fe] $< 0.12$~dex in the panels of $10 < \tau < 12$~Gyr.   The lower panels of age lower than 10~Gyr show the dominant population of the low-[$\alpha$/Fe] and kinematically colder thin disc stars. The lower panels also reveal a small population of kinematically hot, young stars, and as high [$\alpha$/Fe] as the thick disc population in Fig. \ref{fig:afe-feh-age}. To understand their origin, we compare this population of stars with the thick disc stars (high [$\alpha$/Fe] and old), and we discuss the results in Section \ref{younghal}.

\begin{figure*}
	\centering
	\includegraphics[width=\textwidth]{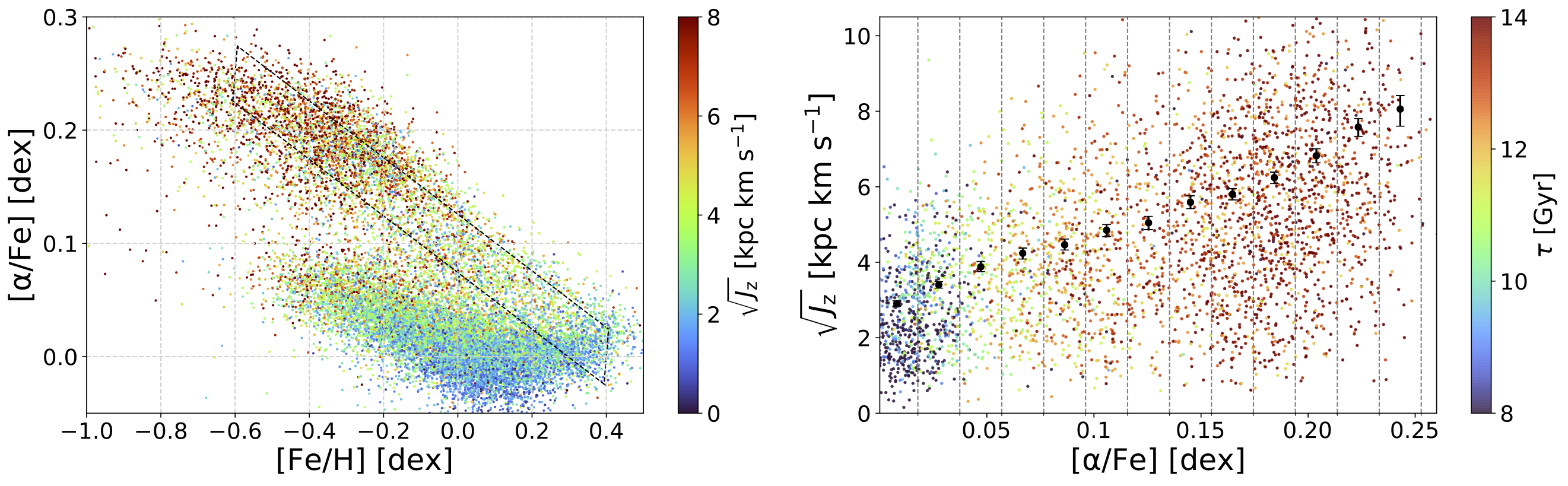}
	\caption{Left panel: the distribution of [$\alpha$/Fe] and [Fe/H] coloured by the square root of the vertical action, $\sqrt{J_z}$. The ridge selection in the left panel represents the highest metallicity track in the [Fe/H]-[$\alpha$/Fe] space. Right panel: the  [$\alpha$/Fe]-$\sqrt{J_z}$ relationship coloured by age in the ridge region highlighted in the left panel. We overlay the scale heights (black dots) and uncertainties (error bars) measured by fitting an isothermal profile to the distribution of $p(J_z)$ in 13 bins in [$\alpha$/Fe]. Analysis of the high-[Fe/H] ridge shows that $J_z$ smoothly decreases with [$\alpha$/Fe] and age.}
	\label{fig:age-jz}
\end{figure*}

\begin{figure*}
	\centering
	\includegraphics[width=\textwidth]{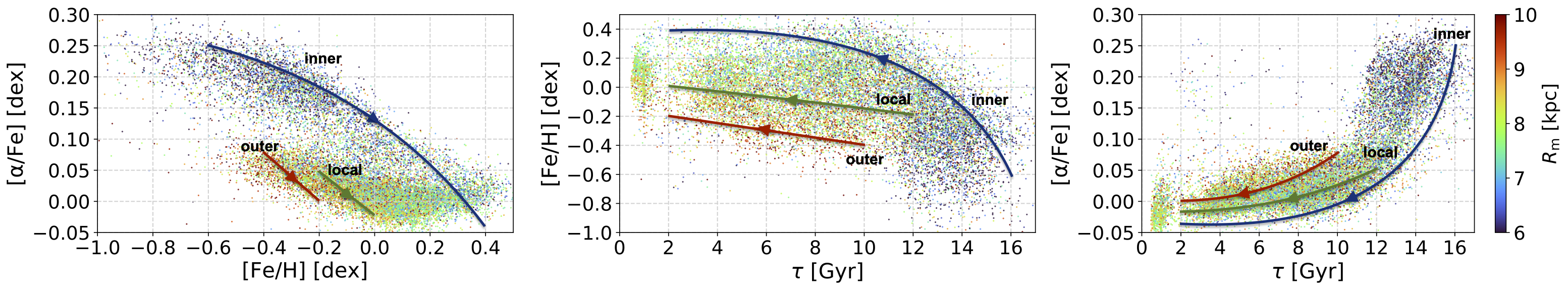}
	\caption{The distribution in [$\alpha$/Fe] vs [Fe/H] (left panel), [Fe/H] vs age (middle panel), and [$\alpha$/Fe] vs age (right panel) coloured by mean orbital radius, $R_{\rm m}$. The ``inner", ``local" and ``outer" arrows indicate the schematic chemical evolution paths at the inner ($R_{\rm m}$ $\sim 6$~kpc), local, i.e. solar radius ($R_{\rm m}$ $\sim 8$~kpc), and outer discs ($R_{\rm m}$ $\sim 10$~kpc), respectively. The metal-poor, outer disc stars follow a different chemical evolution pathway than the inner disc. These evolutionary paths are shown to describe qualitative trends of the chemical evolution at the different radii of the Galactic disc, and are not meant to indicate the chemical evolution paths quantitatively.}
	\label{fig:rm-plot}
\end{figure*}

To further examine whether or not there is a clear distinction between the high and metal-rich, low-[$\alpha$/Fe] sequence, we select the high-metallicity ridge shown in the left panel of Fig. \ref{fig:age-jz}. The ridge is considered to represent the most advanced chemical evolution path of the stars born in the inner disc, $R<6$~kpc, \citep[e.g.,][]{Scho_2009}, and in fact, as shown in the left panel of Fig.  \ref{fig:rm-plot}, their mean orbital radius is always smallest among the same [$\alpha$/Fe] population. For the stars within this high-metallicity ridge, we divided the samples according to their [$\alpha$/Fe]. We measure the scale height using a Markov-Chain Monte Carlo approach, where we fit the $J_z$ distribution in each [$\alpha$/Fe] bin with an isothermal profile, i.e., $p(J_z) \sim \exp{(-J_z/h_{J_z})}$ \citep{Binney_2010, BinneyMcMillan_2011, Ting_2019}. We compute the scale height,  $h_{J_z}$, and its uncertainty in 13 selected bins using [$\alpha$/Fe]. The results, shown in the right panel of Fig. \ref{fig:age-jz}, reveal a smooth decrease of $J_z$ with decreasing [$\alpha$/Fe] and age indicated by colour. The derived $h_{J_z}$ for the different [$\alpha$/Fe] bins also show a smooth decrease with decreasing [$\alpha$/Fe]. The oldest stars are kinematically hotter and higher [$\alpha$/Fe]. This result is consistent with an upside-down formation of the Milky Way \citep{Brook2012, Bird2013}. Although it is subtle, our results also suggest that the decrease in $h_{J_z}$  with decreasing [$\alpha$/Fe] happens more rapidly for the high-[$\alpha$/Fe], old population than for the young population as can be seen from the changing slope in the right panel of Fig. \ref{fig:age-jz}. The change of the slope happens roughly at [$\alpha$/Fe] $\sim$ 0.12, where the Bridge region starts. Overall, the high-[$\alpha$/Fe] thick disc is smoothly connected to the low-[$\alpha$/Fe] thin disc population. This result indicates that the chemo-dynamical evolution of the inner disc is smooth.

In Fig. \ref{fig:rm-plot}, we show the  relations between [$\alpha$/Fe],  [Fe/H] and age coloured by the mean radius of the stellar orbit, $R_{\rm m}$. From the [$\alpha$/Fe]-[Fe/H] relationship, shown in the left panel, we infer that the high-[Fe/H] ridge region selected in the left panel of Fig.  \ref{fig:age-jz} is mainly populated by small $R_{\rm m}$ stars, which is consistent with our view that this region is tracing the chemical evolution of the inner disc. On the other hand, it is known that the low [Fe/H] ([Fe/H] $\lesssim -0.1$), low-[$\alpha$/Fe] population are not connected with the thick high-[$\alpha$/Fe] population and show a distinct population \citep[e.g., ][]{Hayden_2015, Queiroz_2019}. However, as we discussed in Figs. \ref{fig:afe-feh-age} and \ref{fig:al-feh-jz}, it is connected via the Bridge region. Interestingly, as seen in Fig. \ref{fig:al-feh-jz}, the low-[Fe/H], low-[$\alpha$/Fe] stars only appear at age < 11~Gyr. In addition, their $R_{\rm m}$ is predominantly larger ($R_{\rm m}$ > 9~kpc). Hence, we consider that the low-[Fe/H], low-[$\alpha$/Fe] stars formed at the outer disc and their star formation started when the disc grew large enough to develop a wide range of [Fe/H], i.e. the metallicity gradient, at the end of the transition period of the Bridge after the old thick disc formation. As a result, the star formation and chemical evolution path should be different from the inner disc, and the stars in the outer disc do not originate in the thick disc formation phase. 

This different path of the disc formation in the inner disc and the outer disc is schematically described with the arrows in Fig. \ref{fig:rm-plot}. The arrows highlighted with ``inner", ``local" and ``outer" indicate the chemical evolution paths at the inner, local, i.e. solar radius, and outer discs, respectively, inferred from our data. The middle panel shows that low-[Fe/H] stars start forming later than the inner disc and are systematically younger than the thick disc, which is formed only in the inner disc. The right panel shows that the lower-[Fe/H] outer thin disc stars are higher [$\alpha$/Fe]. This is seen as a positive [$\alpha$/Fe] radial gradient in the thin disc \citep[e.g.,][]{Hayden_2015}.

\subsection{The young high-[$\alpha$/Fe] stars}
\label{younghal}
\begin{figure}
	\centering
	\includegraphics[width=\columnwidth]{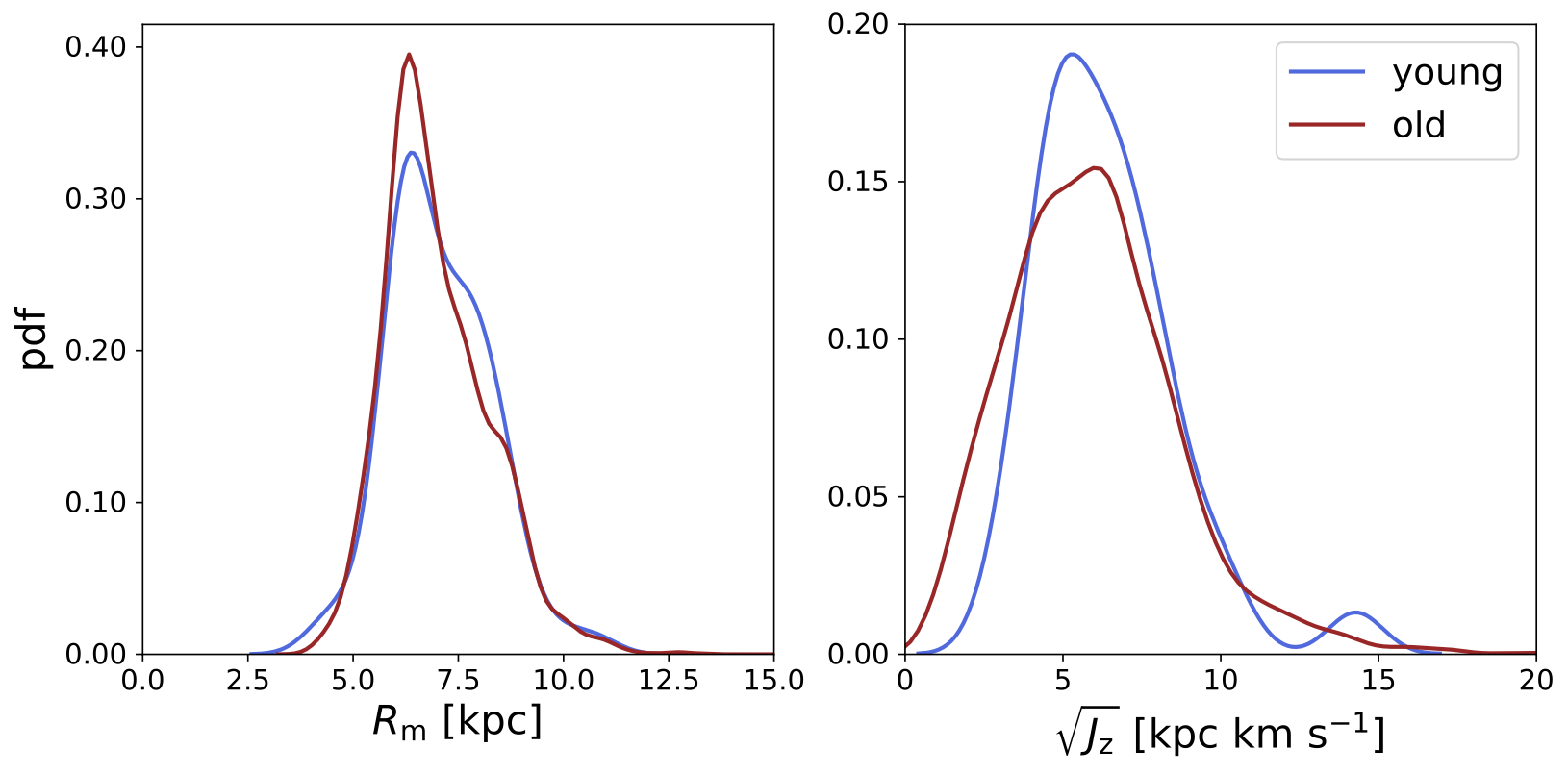}
	\caption{The distribution in $R_{\rm m}$ (left panel) and $\sqrt{J_z}$ (right panel) for old ($\log(\tau {\rm [Gyr]})>1.0$) stars with [$\alpha$/Fe] $>0.12$~dex, shown in red, and young ($\log(\tau {\rm [Gyr]})<0.8$) stars with [$\alpha$/Fe] $>0.12$~dex, shown in blue. The two populations have similar kinematics.}
	\label{fig:hal_hfe_rmean_jz}
\end{figure}

\begin{figure}
	\centering
	\includegraphics[width=\columnwidth]{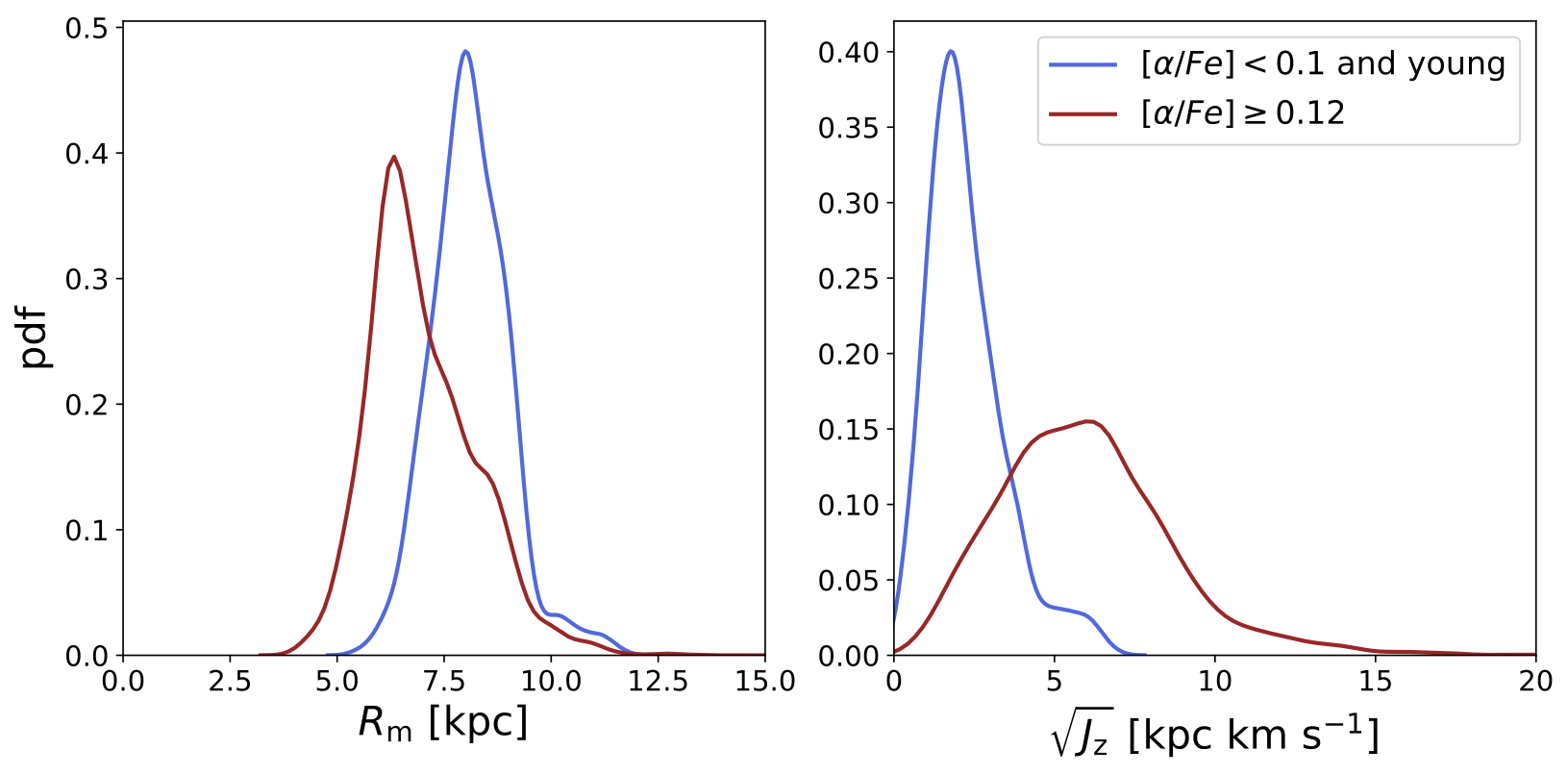}
	\caption{The distribution in $R_{\rm m}$ (left panel) and $\sqrt{J_z}$ (right panel) for stars with [$\alpha$/Fe] $>0.12$~dex, shown in red, and young ($0.2<\log(\tau {\rm [Gyr]})<0.5$) stars with [$\alpha$/Fe] $<0.1$~dex, shown in blue. The kinematical properties differ between the two populations.}
	\label{fig:fig2_hal_hfe_rmean_jz}
\end{figure}

The lower panels of Fig. \ref{fig:al-feh-jz}, consisting of stars younger than 10~Gyr, reveal the existence of a population of kinematically hot, high-[$\alpha$/Fe] stars. To understand their origin, we look at the distribution in $R_{\rm m}$ and $J_z$ between stars with [$\alpha$/Fe] $>0.12$~dex and old ($\log(\tau {\rm [Gyr]})>1.0$) and stars with [$\alpha$/Fe] $>0.12$~dex and young ($\log(\tau {\rm [Gyr]})<0.8$). As shown in Fig. \ref{fig:hal_hfe_rmean_jz} the two groups of stars overlap significantly in both $R_{\rm m}$ and $J_z$ distributions. Fig. \ref{fig:fig2_hal_hfe_rmean_jz}, where we compared between [$\alpha$/Fe] $>0.12$~dex stars and young stars ($0.2<\log(\tau {\rm [Gyr]})<0.5$) having [$\alpha$/Fe] $<0.1$~dex, shows that the two populations differ greatly in their kinematical properties. 

These results indicate that the young high-[$\alpha$/Fe] population originated from the old high-[$\alpha$/Fe], i.e. old thick disc, population rather than the low-[$\alpha$/Fe] thin disc population. Their hot kinematics implies that the high-[$\alpha$/Fe] stars identified as young using the ages given by BINGO have originated from the star-forming gas disc which is as kinematically hot as the gas disc that created the old high-[$\alpha$/Fe] stars. Such a highly turbulent gas disc only existed at the early epoch of the old high-[$\alpha$/Fe] thick disc formation. Therefore, it is likely that these [$\alpha$/Fe] stars identified as young formed at the early epoch, but they are identified as young stars most likely because they are the merger of old binary stars \citep{Jofre_2016}, which has lowered their [C/N] abundance \citep{Izzard_2018} and biased the age estimator. Although we cannot establish when the merger took place, we can conclude from their hot kinematics that the high-[$\alpha$/Fe] stars that appear young originated from the old high-[$\alpha$/Fe] thick disc stars.

By combining the age information with the chemistry and kinematics, we can constrain the origin of the kinematically hot young high-[$\alpha$/Fe] population. Our results are in agreement with \cite{Aguirre_2018}, \cite{Miglio_2020}, who also found similar kinematical properties between young high-[$\alpha$/Fe] stars and young low-[$\alpha$/Fe]. We confirmed their results with a larger number of 69 young high-[$\alpha$/Fe] stars. \cite{Miglio_2020} also found a higher fraction of young (over-massive) high-[$\alpha$/Fe] population in RC stars than RGB stars, and discussed that this is consistent with a scenario that these young [$\alpha$/Fe] stars are the merged binaries, because more binaries are expected to have undergone an interaction around the tip of RGB than fainter RGB.

\section{Implications for the disc formation and evolution}
\label{sec:disc}
Our results suggest a formation scenario for the Galactic disc that involves distinct star formation and chemical evolution pathways of the inner and outer discs. In the inner disc, the thick disc forms early on from chemically well-mixed and turbulent gas, which can be, for example, associated with gas-rich mergers \citep[e.g.,][]{Brook_2004}, cold gas flow accretion \citep[e.g.,][]{Keres_2005,  Dekel_2006, Brooks_2009, Ceverino_2010, Fernandez_2012} or, most likely, a complex interplay of both \citep[e.g.,][]{Grand2018, Grand_2020}. Such a thick disc formation scenario can explain the clear and tight age-[$\alpha$/Fe] (Fig. \ref{fig:age-afe-mh}), age-[Fe/H] (Fig. \ref{fig:age-mh-alphafe}) and [Fe/H]-[$\alpha$/Fe] sequence (Fig. \ref{fig:afe-feh-age}) for the old high-[$\alpha$/Fe] thick disc stars.


After the formation of the old high-[$\alpha$/Fe] thick disc, in the inner region there could be a smooth chemodynamical evolution from high-[$\alpha$/Fe] to low-[$\alpha$/Fe] and increasing metallicity as indicated by the ``inner" pathway in Fig.  \ref{fig:rm-plot}. There is no distinct epoch of thick and thin disc formation, as seen in the ridge region of Fig. \ref{fig:age-jz}. Instead, the thicker to thinner disc transition happens in a smooth manner as stars continue to form with lower $J_z$ from the dense cold gas continuously present at this radius \citep{Brook2012, Grand2018}. The smooth transition between the thick and thin discs in the inner region naturally arises in multi-zone semi-analytical chemo-dynamical evolution models \citep[e.g.,][]{Scho_2009, Sch2017}, where stars keep forming from the left-over gas of the high-[$\alpha$/Fe] sequence. 

A smooth transition between the formation of the thick and thin discs in the inner region is also suggested as the ``centralised starburst pathway" in \cite{Grand2018}. Using the high-resolution \textsc{Auriga} cosmological simulations of the Milky Way \citep{Grand_2017}, \cite{Grand2018} propose the ``centralised starburst pathway" model that can explain the single sequence of the [$\alpha$/Fe]-[Fe/H] distribution in the inner disc seen in the APOGEE data of \cite{Hayden_2015} \citep[see also][]{Palla_2020}. In their model, a major gas-rich merger and cold gas accretion at an early epoch initiates a short period of intense star formation in the inner region during which the thick disc forms with higher [$\alpha$/Fe]. Once Type Ia SNe become significant in chemical enrichment after the peak of star formation in $\sim$ 1~Gyr timescale, more metal-rich low-[$\alpha$/Fe] thin disc stars continuously form from the left-over less turbulent gas in the inner disc. As a result, there is no gap in the formation of the thick and thin disc and a single sequence of [$\alpha$/Fe]-[Fe/H] is expected in the inner disc. Then, we can observe such inner disc stars in our data due to radial migration \citep[e.g.,][]{Brook2012, Minchev_2013, Kawata_2018, Renaud_2020}, which brings them within 7 < $R$ < 9~kpc.


Our results also suggest that the star formation and chemical evolution in the outer disc starts after the thick disc phase. When the thick-disc like, gas-rich merger- and/or cold accretion-dominant, turbulent star formation ends, the galactic halo may have grown enough for the hot gas accretion mode to become dominant \citep{Brooks_2009, Noguchi_2018}. Then, the violent cold gas accretion stops, and the gas disc can grow in an inside-out fashion, as fresh low [Fe/H] gas is accreted smoothly from the hot halo gas. The disc rapidly grows large enough to develop a negative metallicity gradient as seen in the Bridge region of Fig. \ref{fig:afe-feh-age}, unlike for the turbulent small thick disc phase, where the metals are well mixed, and no metallicity gradient can develop. Hence, the metal-poor outer disc developed after the thick disc formation, as indicated by the arrow of the ``outer" disc chemical evolution pathway in Fig. \ref{fig:rm-plot}. 

The Bridge region in Fig. \ref{fig:afe-feh-age} shows that the range of [Fe/H] becomes wider for the younger stars, as indicated by the arrows in the left panel of Fig.~\ref{fig:afe-feh-age}. We consider that the Bridge region is where the thin disc formation begins, and the disc is developing a metallicity gradient with younger stars forming with a broader range of [Fe/H]. Radial migration brings stars formed in the inner disc and outer disc to the solar neighbourhood, which is where the stars in our samples lie \citep[but see also][who claim that radial migration is not very efficient]{Vincenzo_2020, Khoperskov_2020}. As a result, we can observe the mixed chemical distribution from pathways in the inner and outer discs \citep{Scho_2009} . The high metallicity ridge highlighted in Fig. \ref{fig:age-jz} represents the chemical evolution of the inner disc. The low-[Fe/H] and low-[$\alpha$/Fe] stars came from the outer disc. As a result, we observe the two sequences of the high- and low-[$\alpha$/Fe] stars in our sample \citep{Brook2012, Grand2018}. This is consistent with what is seen in the previous observational studies \citep[e.g.][]{Hayden_2017,Feuillet_19}. Our results provide more robust trends with more reliably measured relative ages.

Consequently, the stars currently around the solar radius follow the so-called two-infall model \citep{Grisoni_2017,Spitoni_2019}, although the high-[$\alpha$/Fe] populations are formed in the inner disc and brought to the solar neighbourhood likely by radial migration. On the other hand, if we could select only the stars formed locally, the star formation starts at a later epoch from lower metallicity gas, when the star-forming disc becomes as large as the solar radius. In other words, the stars started forming locally around the solar radius from the second infall of gas, whose metallicity is lower, because of the dilution with the metal-poor gas accreted from the halo \citep{Calura_2009, Spitoni_2019, Buck_2020}. The scenario is also consistent with what is suggested by \cite{Snaith_15} and \cite{Haywood_16,Hay2019}. In fact, by private communication with Misha Haywood after the submission of the first version of this paper, we realised that our schematic view in Fig.~\ref{fig:rm-plot} is consistent with Fig.~6 of \cite{Hay2019}, except that they considered that the outer low-[$\alpha$/Fe] stars formed earlier than the inner low-[$\alpha$/Fe] stars.

\section{Summary}
\label{sec:summary}

In this paper, we determine precise relative stellar age estimates for 17,305 evolved stars in the APOGEE DR14 survey using a Bayesian Neural Network trained on the APOKASC-2 asteroseismic dataset. To minimize the bias in our age inference, we erase the distance dependence of metallicity and age in our training set by randomly displacing the distance of the stars. We also augment the dataset by over-sampling young and very old stars, to obtain a balanced training data and minimize the effect of regression dilution. Using the chemo-kinematical information, we separate the Galactic disc into three components, the thick and thin discs and the Bridge in the [Fe/H]-[$\alpha$/Fe] distribution. The thick disc population is older and higher-[$\alpha$/Fe] ([$\alpha$/Fe] $\gtrsim$ 0.12) than the thin disc. We argue that the Bridge population connects the thick disc and thin disc phases smoothly, rather than being part of the traditional thick disc. We also find an unusual population of young and high-[$\alpha$/Fe] stars. However, we found that their kinematic properties are similar to the old high-[$\alpha$/Fe] stars, which suggests that their origin must be the same as the old high-[$\alpha$/Fe] stars. They are identified as young stars likely due to the merger of binary stars which decreased [C/N] and led to the predicted young ages.

To further investigate whether or not there is a smooth transition between the formation of the thick and thin disc in the inner region, we select a high-metallicity ridge region in the [Fe/H]-[$\alpha$/Fe] plane that follows a continuously increasing [Fe/H] and decreasing [$\alpha$/Fe] sequence. We examined the variation of $J_z$ with  [$\alpha$/Fe] and age and concluded that, while there seems to be a hint of a sudden decrease in $J_z$ around [$\alpha$/Fe] $\sim$ 0.12~dex, $J_z$ smoothly decreases with [$\alpha$/Fe] and also with age. We find that the oldest stars are kinematically hotter and enhanced in $\alpha$-abundances than the younger stars. We found that the high-metallicity ridge is dominated by the stars from the inner disc and traces the continuous chemical evolution of the inner disc, $R<6$~kpc. The formation of the thick disc is expected to happen in a compact disc, i.e. only in the inner disc, and a turbulent period of intense chemical mixing leads to the relatively tight sequence in the distribution of [$\alpha$/Fe] and [Fe/H] for the old stars. From our results, we infer that the inner disc continuously forms stars from the left-over gas after the thick disc formation phase and the subsequent accreting gas, and develop high-[Fe/H] and low-[$\alpha$/Fe] stars. 

We also found that the outer low-[Fe/H] and low-[$\alpha$/Fe] stars are significantly younger than the inner high-[$\alpha$/Fe] ([$\alpha$/Fe] $\gtrsim$ 0.12~dex) stars. We argue that the outer metal-poor disc stars form after the end of cold-mode dominated violent thick disc formation phase. This likely corresponds to the transition from the cold to hot mode of the gas accretion due to the halo mass growth \citep{Noguchi_2018, Grand2018}.

In light of these results, we argue that the inner and outer discs of the Milky Way follow different chemical evolution pathways. After the violent thick disc formation phase ends, the thin disc formation starts with a smaller disc which is as small as the thick disc, and then the thin disc grows in an inside-out fashion. As the disc is growing with a supply of accreting low-[Fe/H] gas, metallicity gradients naturally arise, with the outer disc being more metal-poor than the inner disc. We found that the Bridge region shows a broader range of [Fe/H] with decreasing age, and suggest that the Bridge region is where the thin disc formation begins, and the disc is developing a metallicity gradient. 


The recent work of \cite{Grand_2020} suggested that the last significant merger of Gaia-Enceladus-Sausage \citep[GES,][]{Brook_2003, Belokurov_2018, Haywood_2018a, Helmi_2018} was a gas-rich merger that was essential in forming the thick disc. This picture is also consistent with what we found in this paper because this gas-rich merger can induce a violent starburst in the inner disc due to the dissipation of the gas during the merger. If the GES merger was the last significant merger, then the thin disc phase could start after the GES merger settled. If this scenario is true, the end of the GES merger could correspond to the high-[$\alpha$/Fe] tip ([$\alpha$/Fe] $\sim$ 0.12~dex, [Fe/H] $\sim$ 0.0~dex) of the Bridge region of Fig. \ref{fig:afe-feh-age}. After that, the thin disc grew inside-out from the smooth accretion of the low metallicity gas from the hot halo gas. Although this is admittedly pure speculation, we could test this hypothesis if we measured the relative difference in ages between the GES, the GES merger remnants \citep[e.g.,][]{Chaplin_2019, Montalban_2020}, high-[$\alpha$/Fe] thick disc and the Bridge. Measuring the age difference of stars precisely represents the holy grail of Galactic archaeology, as it allows us to improve our understanding of stellar evolution and probe deeper into the formation and evolution history of the Milky Way.

\section*{Data Availability}

The data underlying this article will be shared on reasonable request to the corresponding author.

\section*{Acknowledgements}
We thank our anonymous referee for their helpful comments to improve the manuscript.
IC and DK acknowledge the support of the UK's Science and Technology Facilities Council (STFC Grant ST/N000811/1 and ST/S000216/1). IC is also grateful for the STFC Doctoral Training Partnerships Grant (ST/N504488/1). IC thanks the LSSTC Data Science Fellowship Program, where their time as a Fellow has benefited this work. AM acknowledges support from the ERC Consolidator Grant funding scheme (project ASTEROCHRONOMETRY, grant agreement number 772293). GRD has received funding from the European Research Council (ERC) under the European Union's Horizon 2020 research and innovation programme (CartographY GA. 804752). This work has made use of data from the European Space Agency (ESA) mission \textit{Gaia} (\url{https://www.cosmos.esa.int/gaia}), processed by the \textit{Gaia} Data Processing and Analysis Consortium (DPAC, \url{https://www.cosmos.esa.int/web/gaia/dpac/consortium}). Funding for the DPAC has been provided by national institutions, in particular, the institutions participating in the \textit{Gaia} Multilateral Agreement. 
This work was inspired from our numerical simulation studies used the UCL facility Grace and the Cambridge Service for Data Driven Discovery (CSD3), part of which is operated by the University of Cambridge Research Computing on behalf of the STFC DiRAC HPC Facility (www.dirac.ac.uk). The DiRAC component of CSD3 was funded by BEIS capital funding via STFC capital grants ST/P002307/1 and ST/R002452/1 and STFC operations grant ST/R00689X/1. DiRAC is part of the National e-Infrastructure.





\bibliographystyle{mnras}
\bibliography{icms}







\bsp	
\label{lastpage}
\end{document}